\documentclass[12pt]{article}
\usepackage{amsmath,amssymb,theorem,cite,epsfig,url,psfrag,eepic,mathtools,amsmath}
\advance \topmargin by -\headheight
\advance \topmargin by -\headsep     
\evensidemargin \oddsidemargin
\def\Maketitle{{\def\newpage{}\maketitle}}
\makeatletter
\def\Appendix{\appendix
  \def\@seccntformat##1{Appendix~\csname the##1\endcsname.~~}}
\makeatother
\makeatletter
\@addtoreset{equation}{section}

\makeatother

\def\XXint#1#2#3{{\setbox0=\hbox{$#1{#2#3}{\int}$}
\vcenter{\hbox{$#2#3$}}\kern-.5\wd0}}

\begin{document}
\title{\textbf{On $W$ algebras commuting with a set of screenings}\vspace*{.3cm}}
\date{}
\author{Alexey Litvinov$^{1,2}$ and Lev Spodyneiko$^{1,3}$\\[\medskipamount]
\parbox[t]{0.85\textwidth}{\normalsize\it\centerline{1. Landau Institute for Theoretical Physics, 142432 Chernogolovka, Russia}}\\
\parbox[t]{0.85\textwidth}{\normalsize\it\centerline{2. Kharkevich Institute for Information Transmission Problems, 127994 Moscow, Russia}}\\
\parbox[t]{0.85\textwidth}{\normalsize\it\centerline{3. California Institute of Technology, 91125 Pasadena, CA, USA}}}
\Maketitle
\begin{abstract}
We consider the problem of classification of all $W$ algebras which commute with a set of exponential screening operators. Assuming that the $W$ algebra has a nontrivial current of spin $3$, we find equations satisfied by the screening operators and classify their solutions.  
\end{abstract}
\section{Introduction}
Consider two-dimensional quantum field theory defined by the action
\begin{equation}\label{action}
   \mathcal{A}=\int\left(\frac{1}{8\pi}g^{ab}\left(\partial_{a}\varphi,\partial_{b}\varphi\right)+\frac{(\mathfrak{r},\varphi)}{4\pi}R+
   \sum_{r=1}^{n}e^{(\mathbf{a}_{r},\varphi)}\right)\,\sqrt{g}\,d^{2}z,
\end{equation}
where $\varphi=(\varphi_{1},\dots,\varphi_{n})$ is the $n$ component bosonic field, $g$ and $R$ are the metric and the scalar curvature on a closed surface and $(\mathfrak{r},\mathbf{a}_{1},\dots,\mathbf{a}_{n})$ is a given set of vectors. We may ask the following question: under what conditions on the set $(\mathfrak{r},\mathbf{a}_{1},\dots,\mathbf{a}_{n})$ the theory defined by the action \eqref{action} has an extended high-spin conformal symmetry (or $W$-symmetry)? 

There is a well known example  of the theory with such a property. It is associated with the root system of the simple Lie algebra $\mathfrak{g}$ of rank $n$. In this case, the vectors $\mathbf{a}_{r}$ are proportional to the simple roots of $\mathfrak{g}$: $\mathbf{a}_{r}=b\boldsymbol{\alpha}_{r}$, and $\mathfrak{r}=b\rho+b^{-1}\rho^{\vee}$ where $\rho$ and $\rho^{\vee}$ are the Weyl vector and the dual Weyl vector correspondingly. The parameter $b$ is  an arbitrary parameter which plays the role of the coupling constant.  The corresponding quantum theory, known as the conformal $\mathfrak{g}-$Toda field theory, possesses extended conformal symmetry generated by the $W(\mathfrak{g})$ algebra with the central charge $c=n+12(\mathfrak{r},\mathfrak{r})$. The full $W(\mathfrak{g})$ algebra contains $n$ independent holomorphic fields with spins equal to the exponents of $\mathfrak{g}$.  The explicit representation of these fields for $A_{n}$, $B_{n}$ and $D_{n}$ series  can be found in \cite{Lukyanov:1990tf}. For $n=1$ the theory \eqref{action} coincides with the celebrated Liouville field theory which plays an important role in quantization of non-critical bosonic string \cite{Polyakov:1981rd}.

It is interesting that there are other $W$-symmetric theories of Toda type \eqref{action} which correspond to Lie superalgebras. There are two well known examples: in rank $n=2$  sine-Liouville theory \cite{FZZ}, and in rank $n=3$ Fateev conformal three-field model \cite{Fateev:1996ea}. Both theories are known to be related to the superalgebras $\mathfrak{sl}(2|1)$ and $D(2|1;\alpha)$ respectively \cite{Feigin:2001yq,Feigin:2004wb}. But unlike the bosonic case the relation is not so obvious. In particular, it is not true that the vectors $\mathbf{a}_{r}$ are proportional to the simple roots of the corresponding superalgebra.  There are also other examples related to superalgebras: the series of theories which corresponds to the superalgebra $\mathfrak{sl}(n|1)$ \cite{Feigin:2004wb} and the series of Toda type theories introduced by Fateev \cite{Fateev:1996em}.

Motivated by the examples listed above it is interesting to find other $W$-symmetric Toda field theories.  The necessary condition for the theory \eqref{action} to have a non-trivial high-spin conformal  symmetry can be formulated as an existence of local chiral fields $W_{s}(z)$ of spin $s>2$ whose modes commute with all the exponential \emph{screening} operators 
\begin{equation}\label{scr-charges}
   \mathcal{S}_{r}\overset{\text{def}}{=}\oint e^{(\mathbf{a}_{r},\varphi(z))}dz,\qquad r=1,\dots,n,
\end{equation}  
i.e. the $W$ algebra is defined as a commutant of a given set of screening operators \cite{zbMATH04093704,Feigin:1990pn,Fateev:1987zh}.  
It is clear that for a generic set of screening operators the commutant is trivial. As we will see below, even the existence of one current of spin greater than $2$ is already too restrictive.  As an example we consider the simplest case: first nontrivial current has spin $s=3$. We find that up to exceptional cases, which we call exotic, the corresponding chiral algebra coincides with the algebra $W(\mathfrak{sl}(m|m'))$  recently introduced in \cite{2015arXiv151208779B}.

This paper is organized as follows. In section \ref{walgebra} we derive equations satisfied by the screening operators which commute with the current of spin $3$. In section \ref{regular-diagrams} we describe the class of regular solutions which corresponds to the  $W(\mathfrak{sl}(m|m'))$ algebras. In section \ref{flip} we discuss the transformation called the flip, which serves as an isomorphism between different regular solutions. In \ref{concl} we give some remarks as well as interesting problems which deserve further studies. In appendices we provide technical details.
\section{W algebra with nontrivial current of spin $3$}\label{walgebra}
Let us formulate the problem in a more precise way. Let $\varphi(z)=(\varphi_{1}(z),\dots,\varphi_{n}(z))$ be the $n-$component holomorphic bosonic field normalized as
\begin{equation}
   \varphi_{i}(z)\varphi_{j}(z')=-\delta_{ij}\log(z-z')+\dots\quad\text{at}\quad z\rightarrow z'.
\end{equation}
Suppose that we fixed the set of vectors $(\mathbf{a}_{1},\dots,\mathbf{a}_{n})$, in general complex,  which form a basis and require that there exists a non-trivial $W$-algebra which commutes with the corresponding set of screening charges \eqref{scr-charges}. By definition it means that there are currents  $W_{s}(z)$ of integer spins $s$ belonging to some set such that
\begin{equation}\label{comm-cond}
  \oint_{\mathcal{C}_{z}}e^{(\mathbf{a}_{r},\varphi(\xi))}W_{s}(z)d\xi=0,
\end{equation}
where $\mathcal{C}_{z}$ is the contour encircling the point $z$. The condition \eqref{comm-cond} should be valid for all $r=1,\dots,n$ and for all $s$ from the set. Moreover, it is natural to assume that the operators $W_{s}(z)$ are descendants of the identity operator, i.e. that $W_{s}(z)$ are differential polynomials of $\partial\varphi(z)$ of degree $s$.

Below we will study the condition \eqref{comm-cond} for $s=2$ and $s=3$. First, we introduce convenient notations. Let $\Gamma_{rs}$ be the (non-degenerate) Gram matrix
\begin{equation}
   \Gamma_{rs}\overset{\text{def}}{=}(\mathbf{a}_{r},\mathbf{a}_{s}),
\end{equation}
and $\hat{\mathbf{a}}_{1},\dots\hat{\mathbf{a}}_{n}$ be the set of vectors orthonormal to $\mathbf{a}_{1},\dots\mathbf{a}_{n}$: 
$(\mathbf{a}_{r},\hat{\mathbf{a}}_{s})=\delta_{rs}$. For convenience we also denote 
\begin{equation}
  (\mathbf{a}_{r},\mathbf{a}_{r})=\Theta_{r}.
\end{equation}

We assume that there exists a current of spin two -- the stress-energy tensor $W_{2}(z)=T(z)$,  which satisfies \eqref{comm-cond}. We take it in the form\footnote{Here and below we assume that all densities are Wick ordered.}
\begin{equation}\label{T}
   T(z)=-\frac{1}{2}(\partial\varphi(z),\partial\varphi(z))+(\mathfrak{r},\partial^{2}\varphi(z)).
\end{equation}
Any exponential field $V_{\boldsymbol{\alpha}}=e^{(\boldsymbol{\alpha},\varphi)}$ (here $\boldsymbol{\alpha}=(\alpha_{1},\dots,\alpha_{n})$)  is a primary field with respect to the stress-energy tensor \eqref{T} with the conformal dimension $\Delta(\boldsymbol{\alpha})=(\boldsymbol{\alpha},2\mathfrak{r}-\boldsymbol{\alpha})/2$. It means that  it has the operator product expansion
\begin{equation}
    T(z)V_{\boldsymbol{\alpha}}(z')=\frac{\Delta(\boldsymbol{\alpha})V_{\boldsymbol{\alpha}}(z')}{(z-z')^{2}}+
    \frac{\partial V_{\boldsymbol{\alpha}}(z')}{z-z'}+\dots,
\end{equation}
where $\dots$ stands for regular terms.  One can easily see that the condition \eqref{comm-cond} is satisfied if and only if $\Delta(\boldsymbol{\alpha})=1$.
So we must have
\begin{equation}\label{1-condition-on-alpha}
   \Delta(\mathbf{a}_{r})=1,\quad\text{for all}\quad r=1,\dots,n.
\end{equation}
Equations \eqref{1-condition-on-alpha} can be considered as a set of conditions on the vector $\mathfrak{r}$. Solving them we find that
\begin{equation}\label{rho}
   \mathfrak{r}=\sum_{r=1}^{n}\left(1+\frac{\Theta_{r}}{2}\right)\hat{\mathbf{a}}_{r}.
\end{equation}
We see that for any non-degenerate system of vectors $\mathbf{a}_{r}$ there exist a current of spin two.

Further, let us assume that there exists a nontrivial current $W_{3}(z)$ of spin $3$. In its most general form it is
\begin{equation}\label{W-most-general}
   W_{3}(z)=C_{ijk}\partial\varphi_{i}\partial\varphi_{j}\partial\varphi_{k}+K_{ij}\partial^{2}\varphi_{i}\partial\varphi_{j}+\Lambda_{j}\partial^{3}\varphi_{j},
\end{equation}
where $C_{ijk}$, $K_{ij}$ and $\Lambda_{j}$ are unknown tensors ($C_{ijk}$ is a totally symmetric tensor). We note that this current is defined up to the transformation
\begin{equation}\label{gauge}
  W_{3}(z)\rightarrow W_{3}(z)+u\, \partial T(z),
\end{equation}
which can be used, for example, to make the tensor $K_{ij}$ traceless. Consider the operator product expansion of the field $W_{3}(z)$ with the exponential field $V_{\boldsymbol{\alpha}}(z')$
\begin{equation}\label{OPE-W}
  W(z)V_{\boldsymbol{\alpha}}(z')=\frac{w(\boldsymbol{\alpha})V_{\boldsymbol{\alpha}}(z')}{(z-z')^{3}}+
  \frac{\Bigl(\eta_{j}(\boldsymbol{\alpha})\partial\varphi_{j}(z)\Bigr)V_{\boldsymbol{\alpha}}(z')}{(z-z')^{2}}+
  \frac{\Bigl(\lambda_{ij}(\boldsymbol{\alpha})\partial\varphi_{i}(z)\partial\varphi_{j}(z)+\nu_{j}(\boldsymbol{\alpha})\partial^{2}\varphi_{j}(z)\Bigr)V_{\boldsymbol{\alpha}}(z')}{(z-z')}+\dots,
\end{equation}
where
\begin{equation}\label{quantities}
\begin{gathered}
  w(\boldsymbol{\alpha})=-C_{ijk}\alpha_{i}\alpha_{j}\alpha_{k}-K_{ij}\alpha_{i}\alpha_{j}-2\Lambda_{i}\alpha_{i},\qquad
  \eta_{i}(\boldsymbol{\alpha})=3C_{ijk}\alpha_{j}\alpha_{k}+K_{ji}\alpha_{j},\\
  \lambda_{ij}(\boldsymbol{\alpha})=-3C_{ijk}\alpha_{k},\qquad
  \nu_{i}(\boldsymbol{\alpha})=-K_{ij}\alpha_{j}.
\end{gathered}
\end{equation}
We must solve the commutativity equation
\begin{equation}
  [\mathcal{S}_{\boldsymbol{\alpha}},W_{3}(z)]\overset{\text{def}}{=}\oint_{\mathcal{C}_{z}}V_{\boldsymbol{\alpha}}(\xi)W_{3}(z)d\xi=0,
\end{equation}
which is equivalent to the condition that the pole in the operator product expansion of $V_{\boldsymbol{\alpha}}(z')$ with $W_{3}(z)$ vanishes. From \eqref{OPE-W} we have
\begin{equation}
 W_{3}(z)V_{\boldsymbol{\alpha}}(z')=\dots+
  \frac{\Bigl(\tilde{\lambda}_{ij}(\boldsymbol{\alpha})\partial\varphi_{i}(z)\partial\varphi_{j}(z)+\tilde{\nu}_{j}(\boldsymbol{\alpha})\partial^{2}\varphi_{j}(z)\Bigr)V_{\boldsymbol{\alpha}}(z)}{(z-z')}+\dots,\quad\text{at}\quad z'\rightarrow z,
\end{equation}
where
\begin{equation}\label{Eqs-on-residue}
  \tilde{\lambda}_{ij}(\boldsymbol{\alpha})=\lambda_{ij}(\boldsymbol{\alpha})-
  \frac{1}{2}\left(\eta_{i}(\boldsymbol{\alpha})\alpha_{j}+\eta_{j}(\boldsymbol{\alpha})\alpha_{i}\right)+\frac{w(\boldsymbol{\alpha})}{2}\alpha_{i}\alpha_{j},\qquad
  \tilde{\nu}_{i}(\boldsymbol{\alpha})=\nu_{i}(\boldsymbol{\alpha})+\frac{w(\boldsymbol{\alpha})}{2}\alpha_{i}.
\end{equation}
So, both  tensors $\tilde{\lambda}_{ij}(\boldsymbol{\alpha})$ and $\tilde{\nu}_{j}(\boldsymbol{\alpha})$ must vanish.

In our case equations \eqref{Eqs-on-residue} must hold for all the vectors $\mathbf{a}_{r}$: $r=1,\dots,n$.  It leads to the system
\begin{equation}\label{Eq1}
   \lambda_{ij}(\mathbf{a}_{r})+\frac{1}{2}\left(\lambda_{ik}(\mathbf{a}_{r})(\mathbf{a}_{r})_{j}(\mathbf{a}_{r})_{k}+\lambda_{jk}(\mathbf{a}_{r})(\mathbf{a}_{r})_{i}(\mathbf{a}_{r})_{k}\right)-
   \frac{1}{2}\left(K_{ki}(\mathbf{a}_{r})_{j}(\mathbf{a}_{r})_{k}+K_{kj}(\mathbf{a}_{r})_{i}(\mathbf{a}_{r})_{k}\right)+
   \frac{w_{r}}{2}(\mathbf{a}_{r})_{i}(\mathbf{a}_{r})_{j}=0
\end{equation}
and 
\begin{equation}\label{Eq2}
  K_{ij}(\mathbf{a}_{r})_{j}=\frac{w_{r}}{2}(\mathbf{a}_{r})_{i},
\end{equation}
where $w_{r}\overset{\text{def}}{=}w(\mathbf{a}_{r})$. The second equation \eqref{Eq2} can be easily solved
\begin{equation}\label{K-sol}
  K_{ij}=\frac{1}{2}\sum_{r=1}^{n}w_{r}(\mathbf{a}_{r})_{i}(\hat{\mathbf{a}}_{r})_{j}.
\end{equation}
The first equation \eqref{Eq1} can be multiplied by $(\mathbf{a}_{s})_{i}(\mathbf{a}_{t})_{j}$ and summed over the indexes $i$ and $j$. As a result we got
\begin{equation}\label{Eq3}
   \mathbb{C}_{rst}+\frac{\Gamma_{rt}}{2}\mathbb{C}_{rrs}+\frac{\Gamma_{rs}}{2}\mathbb{C}_{rrt}-
   \frac{\Gamma_{rs}\Gamma_{rt}}{12}\left(2w_{r}-w_{s}-w_{t}\right)=0.
\end{equation}
where
\begin{equation*}
   \mathbb{C}_{rst}\overset{\text{def}}{=}C_{ijk}(\mathbf{a}_{r})_{i}(\mathbf{a}_{s})_{j}(\mathbf{a}_{t})_{k}.
\end{equation*}
After simple algebra one reduces the equations \eqref{Eq3} to the more convenient form
\begin{equation}\label{Eq4}
    \mathbb{C}_{rrr}\left(1+\Theta_{r}\right)=0,
\end{equation}
and\footnote{Here and below we assume that $\Theta_{r}\neq-2$. One can show that the solution in this case can be obtained as the limit $\Theta_{r}\rightarrow-2$.}
\begin{equation}\label{Eq5}
  \mathbb{C}_{rst}=\frac{\Gamma_{rs}\Gamma_{rt}}{6(\Theta_{r}+2)}\Bigl[(2w_{r}-w_{s}-w_{t})+6\mathbb{C}_{rrr}\Bigr].
\end{equation}

We see that the solution essentially depends on whether $\Theta_{r}=-1$ or not. We shall distinguish these situations and call the vector $\mathbf{a}_{r}$ with $(\mathbf{a}_{r},\mathbf{a}_{r})=\Theta_{r}=-1$ the ``fermionic root'' as opposite to the ``bosonic root''  with $\Theta_{r}\neq-1$. It will be convenient to represent them graphically as:
\begin{equation*}
\begin{picture}(300,30)(200,115)
    \Thicklines
    \unitlength 5pt 
    \put(40,25){\circle{2}}
    \put(42,24.4){-- bosonic root: $\Theta\neq-1$}
    \put(80,25){\circle{2}}
    \put(79.4,24,4){\line(1,1){1.2}}
    \put(79.4,25,6){\line(1,-1){1.2}}
    \put(82,24.4){-- fermionic root: $\Theta=-1$}
  \end{picture}
\end{equation*}
Once the first equation \eqref{Eq4} is resolved we should impose the condition that $\mathbb{C}_{rst}$ is a totally symmetric tensor. The symmetry $s\leftrightarrow t$  is obvious from \eqref{Eq5}, so we have to demand 
\begin{equation}\label{Symm-Eq}
  \mathbb{U}_{rst}\overset{\text{def}}{=}\mathbb{C}_{rst}-\mathbb{C}_{srt}=0.
\end{equation}
Evidently, equation \eqref{Symm-Eq} is satisfied identically when $r=s=t$. Then, there are $2$ independent two-point equations:  $\mathbb{U}_{rss}=\mathbb{U}_{srr}=0$ with $r\neq s$ and $2$ independent three-point equations: $\mathbb{U}_{rst}=\mathbb{U}_{rts}=0$ with $r\neq s \neq t$. Moreover we have the set of conditions \eqref{Eq4}. Altogether we have an overdetermined system which has a solution only if the Gram matrix $\Gamma_{rs}$ obeys some special properties. 

Before going further let us mention an important property of the equations \eqref{Eq4}--\eqref{Eq5}. Namely, we note that \eqref{Eq4}--\eqref{Eq5} enjoy the symmetry (keeping $C_{ijk}$, $K_{ij}$ and $\Lambda_{j}$ unchanged)
\begin{equation}\label{symmetry}
   \mathbf{a}_{r}\rightarrow\mathbf{a}_{r}^{\vee}\overset{\text{def}}{=}\frac{2}{\Theta_{r}}\mathbf{a}_{r},
\end{equation}
applied to any bosonic root. It means that the corresponding $W$-algebra will be the same. So the bosonic roots $\mathbf{a}_{r}$ and $\mathbf{a}_{r}^{\vee}$ always appear in pairs. This is not the case for the fermionic root. 

Solving \eqref{Symm-Eq} is a straightforward exercise \footnote{We will give the details in the purely bosonic case in the appendix \ref{bosonic-arguments} and discuss the generic case  in the appendix \ref{exotic}.}.  The non-reducible solution exists only if the Gram matrix is tridiagonal. Analyzing the two-point equations one can show that the Gram matrix must be built from one of the six elementary two-point blocks:
\begin{equation}\label{blocks}
\begin{picture}(330,80)(85,100)
    \Thicklines
    \unitlength 5pt 
    \put(0,30){$\begin{pmatrix}
    2\varkappa&-\varkappa\\
    -\varkappa&2\varkappa
    \end{pmatrix}$}
    \put(1,25){\circle{2}}
    \put(11,25){\circle{2}}
    \put(2,25){\line(1,0){8}}
    \put(5,20){$1$}
    \put(16,30){$\begin{pmatrix}
    2\varkappa^{-1}&-1\\
    -1&2\varkappa
    \end{pmatrix}$}
    \put(18,25){\circle{2}}
    \put(28,25){\circle{2}}
    \put(19,25){\line(1,0){1}}
    \put(20.4,25){\line(1,0){1}}
    \put(21.8,25){\line(1,0){1}}
    \put(23.2,25){\line(1,0){1}}
    \put(24.6,25){\line(1,0){1}}
    \put(26,25){\line(1,0){1}}
    \put(23,20){$2$}
    \put(34,30){$\begin{pmatrix}
    2\varkappa&-\varkappa\\
    -\varkappa&-1
    \end{pmatrix}$}
    \put(35,25){\circle{2}}
    \put(45,25){\circle{2}}
    \put(44.4,24,4){\line(1,1){1.2}}
    \put(44.4,25,6){\line(1,-1){1.2}}
    \put(44.4,24,4){\line(1,1){1.2}}
    \put(44.4,25,6){\line(1,-1){1.2}}
    \put(36,25){\line(1,0){8}}
    \put(40,20){$3$}
    \put(50,30){$\begin{pmatrix}
    2\varkappa^{-1}&-1\\
    -1&-1
    \end{pmatrix}$}
    \put(52,25){\circle{2}}
    \put(62,25){\circle{2}}
    \put(61.4,24,4){\line(1,1){1.2}}
    \put(61.4,25,6){\line(1,-1){1.2}}
    \put(53,25){\line(1,0){1}}
    \put(54.4,25){\line(1,0){1}}
    \put(55.8,25){\line(1,0){1}}
    \put(57.2,25){\line(1,0){1}}
    \put(58.6,25){\line(1,0){1}}
    \put(60,25){\line(1,0){1}}
    \put(57,20){$4$}
    \put(68,30){$\begin{pmatrix}
    -1&-\varkappa\\
    -\varkappa&-1
    \end{pmatrix}$}
    \put(69,25){\circle{2}}
    \put(79,25){\circle{2}}
    \put(68.4,24,4){\line(1,1){1.2}}
    \put(68.4,25,6){\line(1,-1){1.2}}
    \put(78.4,24,4){\line(1,1){1.2}}
    \put(78.4,25,6){\line(1,-1){1.2}}
    \put(70,25){\line(1,0){8}}
    \put(74,20){$5$}
    \put(85,30){$\begin{pmatrix}
    -1&-1\\
    -1&-1
    \end{pmatrix}$}
    \put(86,25){\circle{2}}
    \put(96,25){\circle{2}}
    \put(85.4,24,4){\line(1,1){1.2}}
    \put(85.4,25,6){\line(1,-1){1.2}}
    \put(95.4,24,4){\line(1,1){1.2}}
    \put(95.4,25,6){\line(1,-1){1.2}}
    \put(87,25){\line(1,0){1}}
    \put(88.4,25){\line(1,0){1}}
    \put(89.8,25){\line(1,0){1}}
    \put(91.2,25){\line(1,0){1}}
    \put(92.6,25){\line(1,0){1}}
    \put(94,25){\line(1,0){1}}
    \put(91,20){$6$}
  \end{picture}
\end{equation}
where we present the Dynkin diagrams as well as the corresponding contributions to the Gram matrix. The parameter $\varkappa$ in \eqref{blocks} is an arbitrary parameter which is not fixed by the two-point equations. In order to glue the blocks \eqref{blocks} into the total Gram matrix one must impose also the three-point equations $\mathbb{U}_{rst}=\mathbb{U}_{rts}=0$. In particular, they will relate parameters $\varkappa$ for different blocks. In principle, any combination of solid and dashed lines is possible. However, using the symmetry transformation \eqref{symmetry} one can change the total Dynkin diagram. For example, it relates the first and the second elementary blocks from \eqref{blocks}, as well as the third and the fourth. Below in section \ref{regular-diagrams} we will consider in details the Dynkin diagrams which consist only of solid lines (up to the transformation \eqref{symmetry}), which we call regular diagrams. The others, which can not be reduced by means of the transformation \eqref{symmetry} to this form, i.e. contain at least one dashed line, will be called exotic diagrams. We will consider the corresponding conformal algebras elsewhere.   
\section{Regular diagrams: $W(\mathfrak{sl}(m|m'))$ algebras}\label{regular-diagrams}
Now we describe in details the solution in the case of  regular graph. It consists of any combination of bosonic and fermionic roots glued by the solid lines (i.e. the Gram matrix is a tridiagonal matrix built from the two-point blocks $1$, $3$ and $5$ from \eqref{blocks} only). The bosonic root contributes to the Gram  matrix as
\begin{equation}\label{bosonic-matrix}
   \Gamma_{\textrm{b}}=
    \begin{pmatrix}
    \dots&-\varkappa&0\\
    -\varkappa&2\varkappa&-\varkappa\\
    0&-\varkappa&\dots
    \end{pmatrix},
\end{equation}
while any fermionic as
\begin{equation}\label{fermionic-matrix}
   \Gamma_{\textrm{f}}=
    \begin{pmatrix}
    \dots&-\varkappa_{1}&0\\
    -\varkappa_{1}&-1&-\varkappa_{2}\\
    0&-\varkappa_{2}&\dots
    \end{pmatrix},
    \quad\text{with}\quad
    \varkappa_{1}+\varkappa_{2}=-1
\end{equation}
It is clear that the Gram matrix built of elementary blocks \eqref{bosonic-matrix} and \eqref{fermionic-matrix} contains only one free parameter. The parameters $w_{r}$ and $\mathbb{C}_{rrr}$ are explicitly given by
\begin{equation}
  \Gamma_{r,r+1}\left(w_{r}-w_{r+1}\right)=\omega,\qquad
  \mathbb{C}_{rrr}=-\frac{1}{6}\left(\frac{1}{\Gamma_{r,r+1}}-\frac{1}{\Gamma_{r,r-1}}\right)\,\omega,
\end{equation}
and $\omega$ is an arbitrary nonzero constant. We note that $\mathbb{C}_{rrr}$ is automatically zero for any bosonic root as it should be. Moreover the solution is unchanged under simultaneous shift $\omega_{r}\rightarrow\omega_{r}+\lambda$. This transformation is equivalent to the gauge symmetry \eqref{gauge}.

If there are only bosonic roots the Gram matrix is
\begin{equation}\label{Dynkin}
\Gamma_{rs}=
 \begin{pmatrix}
    \;2\varkappa  & -\varkappa & 0 & \hdotsfor{2} & 0\\
    -\varkappa &  \;2\varkappa & -\varkappa & \hdotsfor{2} & 0\\
    0 & -\varkappa & \hdotsfor{4}\\
    \hdotsfor{4} & -\varkappa & 0\\
    0 & \hdotsfor{2} & -\varkappa & \;2\varkappa & -\varkappa\\
    0 & \hdotsfor{2} & 0 & -\varkappa & \;2\varkappa
   \end{pmatrix},
\end{equation}
It means that the vectors $\mathbf{a}_{r}$ have the form $\mathbf{a}_{r}=b\boldsymbol{\alpha}_{r}$, where $b=\sqrt{\varkappa}$ and $\boldsymbol{\alpha}_{r}$ are the simple roots of $\mathfrak{sl}(n+1)$. The corresponding QFT \eqref{action} is exactly the $\mathfrak{sl}(n+1)$ conformal Toda field theory mentioned in the introduction. 

If one includes the fermionic roots as well, the corresponding $W$ algebra would correspond to the superalgebra $\mathfrak{sl}(m|m')$ with some $m$, $m'$: $m+m'=n+1$. In order to describe this correspondence more precisely it is important to note that there is a symmetry which acts on the space of Gram matrices defined above (built from elementary blocks \eqref{bosonic-matrix} and \eqref{fermionic-matrix}). It serves as an isomorphism between $W$ algebras corresponding to different Gram matrices.  The isomorphisms between different realizations are provided by the transformation which we call the ``flip''. It acts on a root system as follows
\begin{equation}\label{flip-trans}
   \mathbf{a}_{r}\overset{\text{flip}}{\longrightarrow}-\mathbf{a}_{r},\quad
   \mathbf{a}_{r\pm1}\overset{\text{flip}}{\longrightarrow}\mathbf{a}_{r\pm1}+\mathbf{a}_{r},\quad
   \mathbf{a}_{s}\overset{\text{flip}}{\longrightarrow}\mathbf{a}_{s}\quad\text{if}\quad s\neq r,r\pm1.
\end{equation}
One can easily see that the ``flip'' transformation applied to the bosonic root does not change the Gram matrix  and hence the isomorphism between the two $W$ algebras is tautological. Conversely the ``flip'' applied to the fermionic root gives different Gram matrix, but belonging to the same class: built of bosonic and fermionic blocks \eqref{bosonic-matrix}--\eqref{fermionic-matrix}  only.    In this case the ``flip'' sends the fermionic root $\mathbf{a}_{r}$ to the fermionic root $-\mathbf{a}_{r}$, but changes the ``statistics'' of the neighboring roots $\mathbf{a}_{r\pm1}\rightarrow\mathbf{a}_{r\pm1}+\mathbf{a}_{r}$: it sends a bosonic root to a fermionic one and vise versa. In the section \ref{flip} we will give convincing arguments that this transformation provides an isomorphism between $W$ algebras in different realizations. Here we emphasize that using the ``flip''  transformation one can always transform  the Gram matrix with $k>0$ fermionic roots to the Gram matrix with  just one fermionic root:
\begin{equation}\label{super-Dynkin}
\Gamma_{rs}= 
\begin{pmatrix}
2\varkappa_{1}& -\varkappa_{1} &  0& \hdotsfor{6} \\
-\varkappa_{1}& 2\varkappa_{1}  &-\varkappa_{1}&0 &\hdotsfor{5}\\
0 &-\varkappa_{1}  & \ddots & \ddots& \ddots &\hdotsfor{4}\\
\hdots&  \ddots &  \ddots &2\varkappa_{1}&-\varkappa_{1}&0&\hdotsfor{3}\\
\hdotsfor{2}&0 & -\varkappa_{1}& -1 &  -\varkappa_{2} &0 &\hdotsfor{2} \\
\hdotsfor{3}& 0& -\varkappa_{2} &  2\varkappa_{2} &\ddots & \ddots&\hdots\\
\hdotsfor{4}& \ddots & \ddots &\ddots &-\varkappa_{2}  &0 \\
\hdotsfor{5} & 0 &-\varkappa_{2} &2\varkappa_{2}&-\varkappa_{2}\\
\hdotsfor{6}&0 &-\varkappa_{2}&2\varkappa_{2} 
\end{pmatrix}
\begin{matrix}
\left.
\begin{matrix}
\\
\\
\\
\phantom{a}
\end{matrix}
\right\}m-1
\\
\\
\\
\left.
\begin{matrix}
\\
\\
\\
\phantom{a}
\end{matrix}
\right\}
m'-1
\end{matrix}
\end{equation}
with $\varkappa_{1}+\varkappa_{2}=-1$ and for some $m,m'>0$. The matrix \eqref{super-Dynkin} consists of two blocks of size $m-1$ and $m'-1=n-m$ of $\mathfrak{sl}(n)$ type \eqref{Dynkin} glued by the fermionic root. It corresponds to the graph
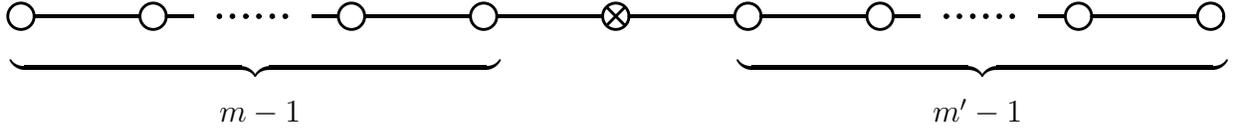
\begin{equation*}
\begin{picture}(300,70)(130,80)
    \Thicklines
    \unitlength 5pt 
    \put(10,25){\circle{2}}
    \put(11,25){\line(1,0){8}}
    \put(20,25){\circle{2}}
    \put(21,25){\line(1,0){2}}
    \put(25,25){\circle{.2}}
    \put(26,25){\circle{.2}}
    \put(27,25){\circle{.2}}
    \put(28,25){\circle{.2}}
    \put(29,25){\circle{.2}}
    \put(30,25){\circle{.2}}
    \put(32,25){\line(1,0){2}}
    \put(35,25){\circle{2}}
    \put(36,25){\line(1,0){8}}
    \put(45,25){\circle{2}}
    \put(55,25){\circle{2}}
    \put(46,25){\line(1,0){8}}
    \put(54.4,24,4){\line(1,1){1.2}}
    \put(54.4,25,6){\line(1,-1){1.2}}
    \put(65,25){\circle{2}}
    \put(56,25){\line(1,0){8}}
    \put(75,25){\circle{2}}
    \put(66,25){\line(1,0){8}}
    \put(76,25){\line(1,0){2}}
    \put(80,25){\circle{.2}}
    \put(81,25){\circle{.2}}
    \put(82,25){\circle{.2}}
    \put(83,25){\circle{.2}}
    \put(84,25){\circle{.2}}
    \put(85,25){\circle{.2}}
    \put(87,25){\line(1,0){2}}
    \put(90,25){\circle{2}}
    \put(91,25){\line(1,0){8}}
    \put(100,25){\circle{2}}
    \put(64.2,22){$\underbrace{\phantom{aaaaaaaaaaaaaaaaaaaaaaaaaaaaaa}}$}
    \put(79,17){$m'-1$}
    \put(9.2,22){$\underbrace{\phantom{aaaaaaaaaaaaaaaaaaaaaaaaaaaaaa}}$}
    \put(25,17){$m-1$}
  \end{picture}
\end{equation*}
We note that the matrix \eqref{super-Dynkin} coincides with the ``dressed'' Cartan matrix for $(\mathfrak{sl}(m|m')$ Lie superalgebra \cite{Feigin:2004wb}. According to \cite{2015arXiv151208779B} the theory defined by the Gram matrix \eqref{super-Dynkin} corresponds to  the $W(\mathfrak{sl}(m|m'))$ algebra. It has the central charge $c(m,m')$ given by
\begin{equation}\label{c(m,n)}
    c(n,m)=\frac{n(n^{2}-1)}{\varkappa_{1}}+\frac{m(m^{2}-1)}{\varkappa_{2}}+(n-m)\bigl((n-m)^{2}-1\bigr)\frac{(\varkappa_{1}-\varkappa_{2})}{2}+
    \frac{1}{2}\left(3(n-m)^{2}-1\right)(n+m)-1,
\end{equation}
with $\varkappa_{1}+\varkappa_{2}=-1$.
\section{The ``flip'' fransformation: fermionic reflection}\label{flip}
In this section we give the  arguments that the ``flip'' transformation \eqref{flip-trans} introduced in the previous section provides an isomorphism between different $W$-algebras. 

Consider the theory defined by the action \eqref{action}. For simplicity we consider the geometry of the two-sphere.
Using the well known trick \cite{Goulian:1990qr} one can show that the $N-$point correlation function
\begin{equation}
  \langle V_{\boldsymbol{\alpha}_{1}}(\xi_{1},\bar{\xi}_{1})\dots V_{\boldsymbol{\alpha}_{N}}(\xi_{N},\bar{\xi}_{N})\rangle,\quad
  \text{where}\quad
  V_{\boldsymbol{\alpha}}=e^{(\boldsymbol{\alpha},\varphi)},
\end{equation}
being considered as a function of the total charge $\boldsymbol{\alpha}=\boldsymbol{\alpha}_{1}+\dots+\boldsymbol{\alpha}_{N}$ has multiple poles at the values
\begin{equation}\label{onshell-cond}
\boldsymbol{\alpha}+\sum_{j=1}^{n}m_{j}\mathbf{a}_{j}=2\mathfrak{r},
\end{equation} 
where $m_{j}$'s are some non-negative integer numbers. The multiple residues at these poles are proportional to the free-field correlation functions
\begin{equation}\label{GL-relation}
  \textrm{Res}\,\langle V_{\boldsymbol{\alpha}_{1}}(\xi_{1},\bar{\xi}_{1})\dots V_{\boldsymbol{\alpha}_{N}}(\xi_{N},\bar{\xi}_{N})\rangle
  \biggl|_{\boldsymbol{\alpha}+\sum m_{j}\mathbf{a}_{j}=2\mathfrak{r}}
  \sim 
  \langle V_{\boldsymbol{\alpha}_{1}}(\xi_{1},\bar{\xi}_{1})\dots V_{\boldsymbol{\alpha}_{N}}(\xi_{N},\bar{\xi}_{N})
  \prod_{j=1}^{n}\frac{\left(\mathcal{S}_{j}\right)^{m_{j}}}{\pi^{m_{j}}m_{j}!}\rangle_{\textrm{\tiny{FF}}},
\end{equation}
where  $\mathcal{S}_{j}=\int e^{(\mathbf{a}_{j},\varphi(\xi,\bar{\xi}))}d^{2}\xi$.  

Now, let us assume that the system $(\mathbf{a}_{1},\dots,\mathbf{a}_{n})$ obeys the rules described in the section \ref{regular-diagrams}: the Gram matrix is a tridiagonal matrix  built of the elementary blocks \eqref{bosonic-matrix},  \eqref{fermionic-matrix} and the vector $\mathfrak{r}$ is given by \eqref{rho}. Suppose that one of the roots is fermionic: say $\mathbf{a}_{r}$ with $(\mathbf{a}_{r},\mathbf{a}_{r})=-1$. Let us pick the contribution of this root to the integral in the r.h.s. of \eqref{GL-relation}. Schematically it has the form
\begin{equation}\label{int1}
   \int\mathcal{D}_{m_{r}}(x)\prod_{i=1}^{m_{r}}\prod_{j=1}^{N+m_{r-1}+m_{r+1}}|x_{i}-t_{j}|^{2p_{j}}\,d^{2}\vec{x}_{m_{r}},
\end{equation}
where
\begin{equation*}
  \mathcal{D}_{n}(x)=\prod_{i<j}|x_{i}-x_{j}|^{2},\qquad
  d^{2}\vec{x}_{n}=\frac{1}{\pi^{n}n!}\prod_{j=1}^{n}d^{2}x_{j}.
\end{equation*}
The set $(t_{1},t_{2},\dots,)$  in \eqref{int1} encodes the coordinates of all the fields which interact with the field $\mathcal{S}_{r}$ and $(p_{1},p_{2},\dots)$ the corresponding exponents. We note that it follows from \eqref{rho} and \eqref{onshell-cond} that the exponents obey the condition 
\begin{equation}
  \sum_{j=1}^{N+m_{r-1}+m_{r+1}}p_{j}=-m_{r}-1,
\end{equation}
which guaranties the absence of the singularity at the infinity. One can show that  the integral \eqref{int1} converges in some domain of the parameters.
There is the well known identity for this type of integrals (see for example \cite{Fateev:2007qn})
\begin{multline}\label{Fateev-integral}
 \int\mathcal{D}_{n}(x)\prod_{i=1}^{n}\prod_{j=1}^{n+m+2}|x_{i}-t_{j}|^{2p_{j}}\,d^{2}\vec{x}_{n}=
 \prod_{j=1}^{n+m+2}\gamma(1+p_{j})\prod_{i<j}|t_{i}-t_{j}|^{2+2p_{i}+2p_{j}}
 \times\\\times
 \int\mathcal{D}_{m}(y)\prod_{i=1}^{m}\prod_{j=1}^{n+m+2}|y_{i}-t_{j}|^{-2-2p_{j}}\,d^{2}\vec{y}_{m},
\end{multline}
provided that $\sum p_{j}=-n-1$ and 
\begin{equation*}
  \gamma(x)=\frac{\Gamma(x)}{\Gamma(1-x)}.
\end{equation*}
Applying the identity \eqref{Fateev-integral} to the free-field correlation function \eqref{GL-relation} we find that
\begin{equation}\label{flip-theories}
 \langle V_{\boldsymbol{\alpha}_{1}}(\xi_{1},\bar{\xi}_{1})\dots V_{\boldsymbol{\alpha}_{N}}(\xi_{N},\bar{\xi}_{N})
  \prod_{j=1}^{n}\frac{\left(\mathcal{S}_{j}\right)^{m_{j}}}{\pi^{m_{j}}m_{j}!}\rangle_{\textrm{\tiny{FF}}}=\prod_{j=1}^{N}\mathcal{N}\bigl(\boldsymbol{\alpha}_{j}\bigr)
  \langle V_{\tilde{\boldsymbol{\alpha}}_{1}}(\xi_{1},\bar{\xi}_{1})\dots V_{\tilde{\boldsymbol{\alpha}}_{N}}(\xi_{N},\bar{\xi}_{N})
  \prod_{j=1}^{n}\frac{\left(\tilde{\mathcal{S}_{j}}\right)^{\tilde{m}_{j}}}{\pi^{\tilde{m}_{j}}\tilde{m}_{j}!}\rangle_{\textrm{\tiny{FF}}},
\end{equation}
where $\mathcal{N}\bigl(\boldsymbol{\alpha}_{j}\bigr)$ are certain normalization factors,  $\tilde{\boldsymbol{\alpha}}_{k}=\boldsymbol{\alpha}_{k}+\mathbf{a}_{r}$ and $\tilde{\mathcal{S}}_{j}=\int e^{(\tilde{\mathbf{a}}_{j},\varphi(\xi,\bar{\xi}))}d^{2}\xi$ with
\begin{equation}\label{flip-trans2}
   \tilde{\mathbf{a}}_{r}=-\mathbf{a}_{r},\quad\tilde{\mathbf{a}}_{r\pm1}=\mathbf{a}_{r\pm1}+\mathbf{a}_{r},\qquad 
   \tilde{\mathbf{a}}_{j}=\mathbf{a}_{j}\quad\text{if}\quad j\neq r ,r \pm1.
\end{equation}
The parameters $m_{j}$ are all unchanged except $m_{r}$: $m_{r}\rightarrow \tilde{m}_{r}$ with
\begin{equation}
  \tilde{m}_{r}=N+m_{r-1}+m_{r-1}-m_{r}-2.
\end{equation}
We note that \eqref{flip-trans2} is exactly the ``flip'' transformation \eqref{flip-trans}. Equation \eqref{flip-theories} should be understood as a map between the two free-field correlation functions. It is natural to assume that this relation holds not only for the residues \eqref{GL-relation}, but also for the total correlation functions. In this case we have an exact correspondence between the two conformal field theories. In particular, it implies that the corresponding conformal algebras are isomorphic.

Let us stress that in discussions above \eqref{flip-theories} we implicitly assumed that all the charges $\boldsymbol{\alpha}_{j}$ are generic. In particular, we assumed that 
\begin{equation}\label{ort-cond}
 (\boldsymbol{\alpha}_{j},\mathbf{a}_{r})\neq0\quad\text{for all}\quad j.
\end{equation}
If one of the inequalities \eqref{ort-cond} is violated we must use the integral identity \eqref{Fateev-integral} with additional care. Indeed, in this case one of the parameters $p_{j}=0$, so the corresponding coordinate does not enter in the l.h.s of the relation  \eqref{Fateev-integral}. In the r.h.s. one has an  indeterminate form $0\times\infty$ with $0$ coming from $\gamma(1+p_{j})$ and $\infty$ coming  from the divergent integral.  This indeterminate form can be easily resolved with the expected result: the coordinate $t_{j}$ simply disappears from \eqref{Fateev-integral}. We note that similar problem exists for any integer value of the scalar product $(\boldsymbol{\alpha}_{j},\mathbf{a}_{r})$ as well as $(\mathbf{a}_{r\pm1},\mathbf{a}_{r})$. So these values must be avoided if possible.

Here we come to an important observation. Let us assume that one of the neighboring roots is bosonic, say  $\mathbf{a}_{r+1}$: $\Theta_{r+1}\neq-1$. As we know bosonic roots always come in pairs: $\mathbf{a}_{r+1}$ and $\mathbf{a}_{r+1}^{\vee}=\frac{2}{\Theta_{r+1}}\mathbf{a}_{r+1}$, so that $(\mathbf{a}_{r+1},\mathbf{a}_{r})=-\Theta_{r+1}/2$ and $(\mathbf{a}_{r+1}^{\vee},\mathbf{a}_{r})=-1$. If we take the root $\mathbf{a}_{r+1}^{\vee}$ in \eqref{GL-relation} instead of $\mathbf{a}_{r+1}$ we will meet exactly the situation described above: there will be integer exponents 
\begin{equation}
  p_{j}=-(\mathbf{a}_{r},\mathbf{a}_{r+1}^{\vee})=1,  
\end{equation}
in the l.h.s. of \eqref{Fateev-integral} and hence we have an indeterminate form. Carefully treating with this singularity we come to the interesting conclusion that the dual screening operator $\oint\exp((\mathbf{a}_{r+1}^{\vee},\varphi(z)))dz$ maps to the screening operator dressed by a differential polynomial of degree $1$. Namely,
\begin{equation}\label{dressed-screening}
    \oint e^{(\mathbf{a}_{r+1}^{\vee},\varphi(z))}dz\overset{\textrm{flip}_{r}}{\longrightarrow}
    \oint (\mathbf{a}_{r},\partial\varphi(z))e^{(\mathbf{a}_{r+1}^{\vee},\varphi(z))}dz.
\end{equation}
We note that the operator $\oint e^{(\mathbf{a}_{r+1},\varphi(z))}dz$ still maps to the exponential operator
\begin{equation}
    \oint e^{(\mathbf{a}_{r+1},\varphi(z))}dz\overset{\textrm{flip}_{r}}{\longrightarrow}
    \oint e^{(\mathbf{a}_{r+1}+\mathbf{a}_{r},\varphi(z))}dz.
\end{equation}

This phenomenon, the appearance of the screening fields dressed by a polynomial \eqref{dressed-screening}, is a special property of systems involving fermionic screenings.  It holds for all theories, not necessarily with tridiagonal Gram matrices. It can be formulated as follows. Suppose that we have a $W$ algebra which commutes with two interacting fermionic screening operators
\begin{equation}\label{2fer-scr}
    \mathcal{S}_{1}=\oint e^{(\mathbf{a}_{1},\varphi)}dz,\quad
    \mathcal{S}_{2}=\oint e^{(\mathbf{a}_{2},\varphi)}dz\qquad\text{with}\quad
    (\mathbf{a}_{1},\mathbf{a}_{1})=(\mathbf{a}_{2},\mathbf{a}_{2})=-1\quad\text{and}\quad
    (\mathbf{a}_{1},\mathbf{a}_{2})\neq0,
\end{equation}
then it also commutes with the dressed screening field
\begin{equation}\label{sm-screening}
    \mathcal{S}_{12}=\oint (\mathbf{a_{1}},\partial\varphi)e^{(\mathbf{b}_{12},\varphi)}dz,\quad\text{where}\quad
    \mathbf{b}_{12}=\frac{2}{(\mathbf{a}_{1}+\mathbf{a}_{2})^{2}}(\mathbf{a}_{1}+\mathbf{a}_{2}).
\end{equation}
We note that the integrand in \eqref{sm-screening} is defined up to the total derivative. Adding the total derivative is equivalent to the shift $\mathbf{a}_{1}\rightarrow\mathbf{a}_{1}+\lambda\mathbf{a}_{2}$. However the choice of the vector $\mathbf{a}_{1}$ (or $\mathbf{a}_{2}$) in the pre exponent in \eqref{sm-screening} is distinguishable. Namely the field  $\mathcal{V}_{12}=(\mathbf{a_{1}},\partial\varphi)e^{(\mathbf{b}_{12},\varphi)}$ behaves as an exponential operator
\begin{equation}
  \mathcal{V}_{12}(z)\mathcal{V}_{12}(z')=(z-z')^{-(\mathbf{b}_{12},\mathbf{b}_{12})}:\mathcal{V}_{12}(z)\mathcal{V}_{12}(z'):.
\end{equation}
\section{Concluding remarks}\label{concl}
In these notes we studied $W$-algebras commuting with the set of exponential screening operators. We found that if the $W$-algebra has a non-trivial current of spin $3$ the Gram matrix of screening charges must be a tridiagonal matrix built of elementary blocks \eqref{blocks}. In the special case which we call regular the corresponding algebra coincides with the algebra $W(\mathfrak{sl}(n|m))$ recently introduced by Bershtein, Feigin and Merzon in \cite{2015arXiv151208779B}.
Here we formulate important conclusions.
\begin{enumerate}
\item As we mentioned in the introduction one can study the same problem as we have studied here, but assuming that the first nontrivial current
$W_{s}$ has spin $s>3$. We studied the next case: $s=4$. The details of our analysis will be presented elsewhere.  Here we formulate the result. Similar to $s=3$ case there are two types of roots: bosonic and fermionic ones.  In the purely bosonic case one finds that the vectors $\mathbf{a}_{r}$ must be proportional to the simple roots of either $A$, $B$, $C$ or $D$-series. If one includes the fermionic roots as well  there will be more solutions. For example, there is a series of solutions for any $n\geq3$ which consists only on fermionic roots. Their Gram matrix corresponds to the following graph
\begin{equation}\label{Dn}
\begin{picture}(300,100)(220,80)
    \Thicklines
    \unitlength 5pt 
    \put(48,32){\circle{2}}
    \put(48,18){\circle{2}}
    \put(54.4,24,4){\line(-1,-1){7}}
    \put(54.4,25,6){\line(-1,1){7}}
    \put(47.4,31.4){\line(1,1){1.2}}
    \put(47.4,18.6){\line(1,-1){1.2}}
    \put(48,19){\line(0,1){12}}
    \put(55,25){\circle{2}}
    \put(54.4,24,4){\line(1,1){1.2}}
    \put(54.4,25,6){\line(1,-1){1.2}}
    \put(66,25){\line(1,0){8}}
    \put(56,25){\line(1,0){8}}
    \put(65,25){\circle{2}}
    \put(64.4,24,4){\line(1,1){1.2}}
    \put(64.4,25,6){\line(1,-1){1.2}}
    \put(75,25){\circle{2}}
    \put(74.4,24,4){\line(1,1){1.2}}
    \put(74.4,25,6){\line(1,-1){1.2}}
    \put(76,25){\line(1,0){2}}
    \put(80,25){\circle{.2}}
    \put(81,25){\circle{.2}}
    \put(82,25){\circle{.2}}
    \put(83,25){\circle{.2}}
    \put(84,25){\circle{.2}}
    \put(85,25){\circle{.2}}
    \put(87,25){\line(1,0){2}}
    \put(90,25){\circle{2}}
    \put(89.4,24,4){\line(1,1){1.2}}
    \put(89.4,25,6){\line(1,-1){1.2}}
    \put(91,25){\line(1,0){8}}
    \put(100,25){\circle{2}}
    \put(99.4,24,4){\line(1,1){1.2}}
    \put(99.4,25,6){\line(1,-1){1.2}}
    \put(64.5,22){$\underbrace{\phantom{aaaaaaaaaaaaaaaaaaaaaaaaaaaaa}}$}
    \put(79.5,17){$n-3$}
    \put(51.5,20){$\varkappa$}
    \put(51.5,29){$\varkappa$}
    \put(38.5,24){$-1-2\varkappa$}
    \put(55.5,26){$-1-\varkappa$}
    \put(69,26){$\varkappa$}
  \end{picture}
\end{equation}
Here an edge with the label $\varkappa$ on it corresponds to the matrix
\begin{equation*}
    \begin{pmatrix}
       -1&-\varkappa\\
       -\varkappa&-1
    \end{pmatrix}
\end{equation*}
The conformal field theory associated with the graph \eqref{Dn} has the central charge
\begin{equation}\label{c-ON}
   c(n)=\frac{n(x-n+2)(2x-n+1)}{2x(x+1)},\quad\text{where}\quad
   x=\begin{cases}
      \varkappa\quad\text{for}\quad n\in2\mathbb{Z}+1\\
      -1-\varkappa\quad\text{for}\quad n\in2\mathbb{Z}
   \end{cases}
\end{equation}
The corresponding conformal algebra is generated by the stress-energy tensor and the spin $4$ field.  The theory defined by the graph \eqref{Dn} for $n=3$ corresponds to the special case of the Fateev conformal three-field model \cite{Fateev:1996ea}. We believe that the CFT with the central charge \eqref{c-ON} is a natural generalization of the model  \cite{Fateev:1996ea} for $n>3$, but unlike the model \cite{Fateev:1996ea} it has only one free parameter. We believe that the conformal model \eqref{Dn} as well as its integrable deformation deserves further studies. We return to it in next publication.

One can go further and study $W$-algebras whose first non-trivial current has spin $5$. In this case one finds that there is a new type of fermionic roots (or better to say parafermionic roots). They have special quantized lengths
\begin{equation}
   \Theta=-3\quad\text{and}\quad\Theta=-\frac{2}{3}.
\end{equation}
Increasing the spin of the first non-trivial current one will find more roots of special length. We found that for any odd spin new roots of quantized length appear. It makes the classification problem more involved.
\item There is another well known way of constructing $W$ algebras. Namely, through the Goddart-Kent-Olive coset construction \cite{Goddard:1986ee}. It is well known that $W(\mathfrak{sl}(n))$ algebra corresponds to the diagonal coset
\begin{equation}
    W(\mathfrak{sl}(n))\sim\frac{\widehat{\mathfrak{sl}}(n)_{k}\times \widehat{\mathfrak{sl}}(n)_{1}}{\widehat{\mathfrak{sl}}(n)_{k+1}}.
\end{equation}
One can expect that the general $W(\mathfrak{sl}(n|m))$ algebra also admits similar representation.  The explicit expression for the central charge \eqref{c(m,n)} suggests  that there are at least two other cases in which this relation might hold:
\begin{equation}\label{corr-coset}
  W(\mathfrak{sl}(n|n))\sim\frac{\widehat{\mathfrak{sl}}(n)_{k}\times \widehat{\mathfrak{sl}}(n)_{-1}}{\widehat{\mathfrak{sl}}(n)_{k-1}},\qquad
  W(\mathfrak{sl}(n|n-1))\sim\frac{\widehat{\mathfrak{sl}}(n)_{k}}{\widehat{\mathfrak{sl}}(n-1)_{k}\times\widehat{\mathfrak{u}}(1)}.
\end{equation}  
It would be interesting to prove relations \eqref{corr-coset} starting form the GKO construction and also to find the coset representation for the general $W(\mathfrak{sl}(n|m))$ theory.

We note that the formula for the central charge  \eqref{c-ON} for the theory \eqref{Dn} suggests that it also might admit the  coset representation
\begin{equation}\label{On-coset}
   \frac{\widehat{\mathfrak{so}}(n+1)_{k}}{\widehat{\mathfrak{so}}(n)_{k}},\quad\text{where}\quad k=x-n+2.
\end{equation}
\item There is a part of the  symmetry generated by the $W(\mathfrak{sl}(n|m))$ algebra which survives an integrable perturbation. One can see it as follows. By definition,  the algebra $W(\mathfrak{sl}(n|m))$ is a commutant of a set of exponential screening fields 
\begin{equation}\label{scr-charges-II}
   \mathcal{S}_{r}\overset{\text{def}}{=}\oint e^{(\mathbf{a}_{r},\varphi(z))}dz,\qquad r=1,\dots,n+m-1,
\end{equation}  
with the Gram matrix $\Gamma_{rs}=(\mathbf{a}_{r},\mathbf{a}_{s})$ given by \eqref{super-Dynkin}. One can check that there is another field $e^{(\mathbf{a}_{0},\varphi(z))}$ with $\mathbf{a}_{0}=-\sum_{r}\mathbf{a}_{r}$ such that
\begin{equation}\label{second-integral}
  \oint_{\mathcal{C}_{z}}e^{(\mathbf{a}_{0},\varphi(\xi))}W_{3}(z)d\xi=\partial V(z),
\end{equation}
for some $V(z)$. It implies that the field $\mathcal{S}_{0}=\oint e^{(\mathbf{a}_{0},\varphi(z))}dz$ commutes with the zero mode of the current $W_{3}(z)$. This fact provides a strong evidence that in the universal enveloping of the   $W(\mathfrak{sl}(n|m))$ algebra there are infinitely many local Integrals of Motion $\mathbf{I}_{k}:$ $k=1,2\dots$, first two of them being
\begin{equation}
   \mathbf{I}_{1}=\frac{1}{2\pi}\int T(z)dz,\qquad \mathbf{I}_{2}=\frac{1}{2\pi}\int W_{3}(z)dz,
\end{equation}
which satisfy the distinguishable property that they commute with the field $\mathcal{S}_{0}$. It can be argued that $\mathbf{I}_{k}$ mutually commute and hence share the same spectrum. It is interesting to consider the diagonalization problem  for the system of IM's $\mathbf{I}_{k}$. It was considered in \cite{Bazhanov:1994ft} for Virasoro algebra ($W(\mathfrak{sl}(2))$)  and in \cite{Litvinov:2013zda} for $W(\mathfrak{sl}(n))$ algebra, where the system of Bethe anzatz equations for the spectrum was found. We studied the general case and found that the same equations as in \cite{Litvinov:2013zda}
work for the generic $W(\mathfrak{sl}(n|m))$ case with mild modification. The results will be published elsewhere.   
\item 
In section \ref{flip} we found that if the $W$ algebra commutes with two fermionic screening operators \eqref{2fer-scr} with non-trivial scalar product then it also commutes with the dressed operator \eqref{sm-screening}. It is clear that \eqref{sm-screening} is a chiral part of the field of the form
\begin{equation}\label{SM-field}
   V=\int (\mathbf{a},\partial\varphi)(\bar{\mathbf{a}},\bar{\partial}\varphi)e^{(\mathbf{b},\varphi)} d^{2}z.
\end{equation}
Taking the field \eqref{SM-field} as a perturbation will change the kinetic term in the action and clearly corresponds to the sigma-model type theory. This phenomenon, the appearance of dressed screening operators,  leads to the interesting duality between two conformal models. One is the Toda type theory \eqref{action} and another is the sigma-model. To illustrate it, consider for example the $W$ algebra defined by two fermionic screenings which corresponds to $W(\mathfrak{sl}(2|1))$ algebra\footnote{This example was first considered by Vladimir Fateev (unpublished).}. We have two vectors $\mathbf{a}_{1}$ and $\mathbf{a}_{2}$ such that
$(\mathbf{a}_{1},\mathbf{a}_{1})=(\mathbf{a}_{2},\mathbf{a}_{2})=-1$ and $(\mathbf{a}_{1},\mathbf{a}_{2})\neq0$.  It is convenient to choose the coordinates such that
\begin{equation}
   \mathbf{a}_{1}=(ia,b),\qquad
   \mathbf{a}_{2}=(-ia,b)\quad\text{with}\quad
   a^{2}-b^{2}=1.
\end{equation}
This system of screening operators corresponds to the sine-Liouville theory (here $\varphi=(\phi,\Phi)$)
\begin{equation}
  \mathcal{A}_{\textrm{SL}}=\int \left(\frac{1}{8\pi}g^{ij}\Bigl(\partial_{i}\phi\partial_{j}\phi+\partial_{i}\Phi\partial_{j}\Phi\Bigr)+e^{b\Phi}\cos(a\phi)
  +\frac{b^{-1}}{8\pi} R\,\Phi\right)\sqrt{g}\,d^{2}z.
\end{equation}
On the other hand the same algebra commutes with the dressed screening field \eqref{sm-screening}.  Taking this field as a perturbation one arrives to the sigma-model action
\begin{equation}
   \mathcal{A}_{\textrm{SM}}=\int
   \left(\frac{1}{8\pi}g^{ij}\partial_{i}\varphi^{\mu}\partial_{j}\varphi^{\nu}G_{\mu\nu}(\varphi)+\frac{1}{4\pi}R D(\varphi)\right)\sqrt{g}\,d^{2}z,
\end{equation}
where the metric $G_{\mu\nu}(\varphi)$ and the Dilaton field $D(\varphi)$ have the form
\begin{equation}
   G_{\mu\nu}(\varphi)=
   \begin{pmatrix}
       1-a^{2}e^{\frac{1}{b}\Phi}&iabe^{\frac{1}{b}\Phi}\\
       iabe^{\frac{1}{b}\Phi}&1+b^{2}e^{\frac{1}{b}\Phi}
   \end{pmatrix},\qquad
   D(\varphi)=\frac{1}{2b}\Phi.
\end{equation}
One can easily show that in the semiclassical limit $b\rightarrow\infty$ this metric is equivalent to the cigar metric \cite{Witten:1991yr}, i.e. we arrive to the duality between the sine-Liouville theory and the cigar CFT known also as Fateev-Zamolodchikov-Zamolodchikov duality \cite{FZZ}.  It would be interesting to find similar relations for other theories described in this paper.
\end{enumerate}
\section*{Acknowledgments}
These notes would not have been possible without the numerous explanations of Misha Bershtein. We also thank Borya Feigin, Misha Lashkevich and Slava Pugai for their interest and useful suggestions. A.L. especially acknowledges Vladimir Fateev for numerous disscusions.  The research of A.L. is supported by RFBR under the grant 15-32-20974.  The research of L.S. is supported by Russian Science Foundation (project No. 14-12-01383).
\Appendix
\section{Solution of \eqref{Symm-Eq} in the bosonic case}\label{bosonic-arguments}
In this appendix we consider in details solution of \eqref{Symm-Eq} in  the case of all bosonic roots: $\Theta_{r}\neq-1$ for all $r=1,\dots,n$. In this case $\mathbb{C}_{rrr}=0$.  First, let us consider the case where the indexes $r$, $s$ and $t$ in \eqref{Symm-Eq} take only two values (say $1$ and $2$). There are only two independent equations in this case $\mathbb{U}_{122}=0$ and $\mathbb{U}_{211}=0$. Explicitly they read 
\begin{equation}\label{Eq8}
 \begin{aligned}
   &\frac{\Gamma_{12}\Theta_{1}}{\Theta_{1}+2}(w_{1}-w_{2})=-\frac{2(\Gamma_{12})^{2}}{\Theta_{2}+2}(w_{1}-w_{2}),\\
   &\frac{\Gamma_{12}\Theta_{2}}{\Theta_{2}+2}(w_{1}-w_{2})=-\frac{2(\Gamma_{12})^{2}}{\Theta_{1}+2}(w_{1}-w_{2}).
 \end{aligned} 
\end{equation}
There are four solutions to \eqref{Eq8}
\begin{enumerate}
\item $w_{1}=w_{2}$ and arbitrary $\Gamma_{12}$, $\Theta_{1}$ and $\Theta_{2}$,
\item $\Gamma_{12}=0$ and arbitrary $\Theta_{1}$, $\Theta_{2}$, $w_{1}$ and $w_{2}$,
\item $\Gamma_{12}\neq0$, $w_{1}\neq w_{2}$ and $\Theta_{2}=\Theta_{1}$, $\Gamma_{12}=-\frac{\Theta_{1}}{2}$,
\item $\Gamma_{12}\neq0$, $w_{1}\neq w_{2}$ and $\Theta_{2}=\frac{4}{\Theta_{1}}$, $\Gamma_{12}=-1$.
\end{enumerate}
We note that the cases $3$ and $4$ are related by the transformation \eqref{symmetry} and hence the solutions of the type $4$ might be droped. Below we will show that the solutions of the type $1$ lead to the degenerate Gram matrices and hence should be dropped as well. The remaining cases $2$ and $3$ can be represented graphically as
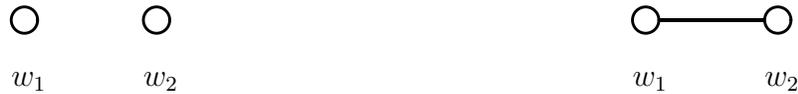
\begin{equation*}
\begin{picture}(300,60)(100,40)
    \Thicklines
    \unitlength 5pt 
    \put(20,15){\circle{2}}
    \put(19,10){$w_{1}$}
    \put(30,15){\circle{2}}
    \put(29,10){$w_{2}$}
    \put(67,15){\circle{2}}
    \put(66,10){$w_{1}$}
    \put(77,15){\circle{2}}
    \put(76,10){$w_{2}$}
    \put(68,15){\line(1,0){8}}
  \end{picture}
\end{equation*}
We note that so far we have no restrictions on the parameters $w_{r}$. They will come out after imposing  the conditions \eqref{Symm-Eq} for three different indexes.

Second, it is easy to show that the vectors with equal $w_{r}$ could not form a connected graph. For example let us consider the case when there are only two non-orthogonal vectors (say $\mathbf{a}_{1}$ and $\mathbf{a}_{2}$) such that the corresponding $w$'s coincide: $w_{1}=w_{2}$ and $\Gamma_{12}\neq0$.  Since the total graph should be connected (the Gram matrix is not a block diagonal matrix) there should exist a vector (say $\mathbf{a}_{3}$) such that either $\Gamma_{13}\neq0$ or $\Gamma_{23}\neq0$  and $w_{3}\neq w_{1}=w_{2}$. Then one can solve the equations \eqref{Symm-Eq} with the conditions specified above. It is easy to see that there is a unique solution in this case and  that the $3\times3$ block of the Gram matrix formed by the vectors $\mathbf{a}_{1}$, $\mathbf{a}_{2}$ and $\mathbf{a}_{3}$ is proportional to
\begin{equation*}
   \begin{pmatrix}
      2&2&-1\\
      2&2&-1\\
      -1&-1&2
   \end{pmatrix}.
\end{equation*}
Since this argument is valid for any vector $\mathbf{a}_{r}$ connected with the pair $(\mathbf{a}_{1},\mathbf{a}_{2})$ such that $w_{r}\neq w_{1}=w_{2}$ one immediately concludes that $\mathbf{a}_{1}=\mathbf{a}_{2}$. Similar logic applies to any connected graph of vectors with coinciding $w$'s.  If all $w$'s coincide, all equations \eqref{Symm-Eq} are satisfied, but the solution is trivial and corresponds to the pure gauge $W_{3}(z)\sim\partial T(z)$. Henceforth, we conclude that all $\omega_{r}$ must be pairwise distinct.

Now, consider all equations \eqref{Symm-Eq} for the three points. There are $8$ of them: $2$ for any pair and $2$ for the triple. It is easy to check that up to permutations, we have three different solutions with pairwise distinct  $w$'s:
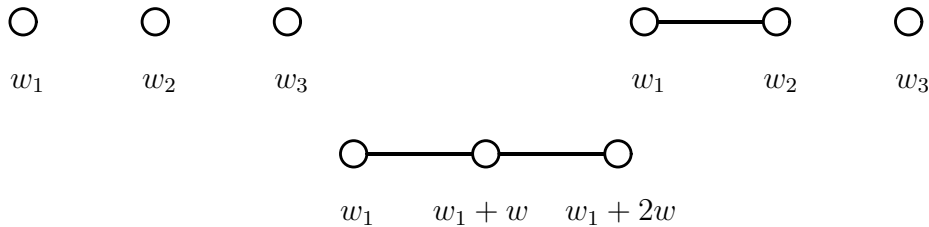
\begin{equation*}
\begin{picture}(300,120)(130,40)
    \Thicklines
    \unitlength 5pt 
    \put(20,25){\circle{2}}
    \put(19,20){$w_{1}$}
    \put(30,25){\circle{2}}
    \put(29,20){$w_{2}$}
    \put(40,25){\circle{2}}
    \put(39,20){$w_{3}$}
    \put(67,25){\circle{2}}
    \put(66,20){$w_{1}$}
    \put(77,25){\circle{2}}
    \put(76,20){$w_{2}$}
    \put(68,25){\line(1,0){8}}
    \put(87,25){\circle{2}}
    \put(86,20){$w_{3}$}
    \put(45,15){\circle{2}}
    \put(44,10){$w_{1}$}
    \put(55,15){\circle{2}}
    \put(51,10){$w_{1}+w$}
    \put(46,15){\line(1,0){8}}
    \put(65,15){\circle{2}}
    \put(56,15){\line(1,0){8}}
    \put(61,10){$w_{1}+2w$}
  \end{picture}
\end{equation*}
where $w$ is an arbitrary parameter. We see that for any triple of connected vectors there is a relation between $w$'s.  Namely, the parameter $\omega_{r}$ homogeneously grows: $\omega_{2}=\omega_{1}+\omega$, $\omega_{3}=\omega_{1}+2\omega$. It is obvious that this relation forbids trivalent vertices and cycles. It means that if the total graph is connected it should be a straight line, i.e. $A_{n}$ graph. Similar logic with mild modifications applies to the case with fermionic roots as well.
\section{Exotic theories}\label{exotic}
In section \ref{regular-diagrams} we described regular theories. The corresponding Gram matrices are built from the two-point blocks $1,3$ and $5$ from \eqref{blocks}.
General solutions also include exotic theories. By definition, the exotic theory is defined by the graph which contains at least one dashed line (up to the transformation \eqref{symmetry}). In order to describe these general solutions it is enough to list all three-point connected graphs with dashed lines. It is easy to see that up to the transformation \eqref{symmetry} and permutations there are $4$ possible three-point graphs
\paragraph{BFF:}
\begin{equation*}
\begin{picture}(300,40)(130,107)
    \Thicklines
    \unitlength 5pt 
    \put(45,25){\circle{2}}
    \put(55,25){\circle{2}}
    \put(46,25){\line(1,0){8}}
    \put(54.4,24,4){\line(1,1){1.2}}
    \put(54.4,25,6){\line(1,-1){1.2}}
    \put(64.4,24,4){\line(1,1){1.2}}
    \put(64.4,25,6){\line(1,-1){1.2}}
    \put(65,25){\circle{2}}
    \put(56,25){\line(1,0){1}}
    \put(57.4,25){\line(1,0){1}}
    \put(58.8,25){\line(1,0){1}}
    \put(60.2,25){\line(1,0){1}}
    \put(61.6,25){\line(1,0){1}}
    \put(63,25){\line(1,0){1}}
  \end{picture}
\end{equation*}
\begin{equation*}
\begin{gathered}
   \Gamma_{123}=
   \begin{pmatrix}
      2\varkappa&-\varkappa&0\\
      -\varkappa&-1&-1\\
      0&-1&-1
   \end{pmatrix},\quad \mathbb{C}_{111}=0\\
  \mathbb{C}_{222}=\frac{2\varkappa+1}{6(\varkappa+1)}(\omega_{1}-\omega_{2}),\quad
  \mathbb{C}_{333}=\frac{1-\varkappa}{6(\varkappa+1)}(\omega_{1}-\omega_{2}),\quad
   \omega_{3}=\omega_{2}+\frac{\varkappa}{\varkappa+1}\left(\omega_{1}-\omega_{2}\right).
  \end{gathered}
\end{equation*}
\paragraph{FBF:}
\begin{equation*}
\begin{picture}(300,40)(130,107)
    \Thicklines
    \unitlength 5pt 
    \put(45,25){\circle{2}}
    \put(55,25){\circle{2}}
    \put(46,25){\line(1,0){8}}
    \put(44.4,24,4){\line(1,1){1.2}}
    \put(44.4,25,6){\line(1,-1){1.2}}
    \put(64.4,24,4){\line(1,1){1.2}}
    \put(64.4,25,6){\line(1,-1){1.2}}
    \put(65,25){\circle{2}}
    \put(56,25){\line(1,0){1}}
    \put(57.4,25){\line(1,0){1}}
    \put(58.8,25){\line(1,0){1}}
    \put(60.2,25){\line(1,0){1}}
    \put(61.6,25){\line(1,0){1}}
    \put(63,25){\line(1,0){1}}
  \end{picture}
\end{equation*}
\begin{equation*}
\begin{gathered}
   \Gamma_{123}=
   \begin{pmatrix}
      -1&-\varkappa&0\\
      -\varkappa&2\varkappa&-1\\
      0&-1&-1
   \end{pmatrix},\quad \mathbb{C}_{111}=-\frac{2\varkappa+1}{6(\varkappa+1)}(\omega_{1}-\omega_{2})\\
  \mathbb{C}_{222}=0,\quad
  \mathbb{C}_{333}=\frac{2+\varkappa}{6(\varkappa+1)}(\omega_{1}-\omega_{2}),\quad
   \omega_{3}=2\omega_{2}-\omega_{1}.
  \end{gathered}
\end{equation*}
\paragraph{FFF:}
\begin{equation*}
\begin{picture}(300,40)(130,107)
    \Thicklines
    \unitlength 5pt 
    \put(45,25){\circle{2}}
    \put(55,25){\circle{2}}
    \put(46,25){\line(1,0){8}}
    \put(44.4,24,4){\line(1,1){1.2}}
    \put(44.4,25,6){\line(1,-1){1.2}}
    \put(54.4,24,4){\line(1,1){1.2}}
    \put(54.4,25,6){\line(1,-1){1.2}}
    \put(64.4,24,4){\line(1,1){1.2}}
    \put(64.4,25,6){\line(1,-1){1.2}}
    \put(65,25){\circle{2}}
    \put(56,25){\line(1,0){1}}
    \put(57.4,25){\line(1,0){1}}
    \put(58.8,25){\line(1,0){1}}
    \put(60.2,25){\line(1,0){1}}
    \put(61.6,25){\line(1,0){1}}
    \put(63,25){\line(1,0){1}}
  \end{picture}
\end{equation*}
\begin{equation*}
\begin{gathered}
   \Gamma_{123}=
   \begin{pmatrix}
      -1&-\varkappa&0\\
      -\varkappa&-1&-1\\
      0&-1&-1
   \end{pmatrix},\quad
  \mathbb{C}_{222}=-\mathbb{C}_{111}=\frac{2\varkappa+1}{6(\varkappa+1)}(\omega_{2}-\omega_{3}),\\
  \mathbb{C}_{333}=\frac{1-\varkappa}{6(\varkappa+1)}(\omega_{1}-\omega_{2}),\quad
   \omega_{3}=\omega_{2}+\frac{\varkappa}{\varkappa+1}(\omega_{1}-\omega_{2}).
  \end{gathered}
\end{equation*}
\paragraph{FFF$'$:}
\begin{equation*}
\begin{picture}(300,40)(130,107)
    \Thicklines
    \unitlength 5pt 
    \put(45,25){\circle{2}}
    \put(55,25){\circle{2}}
    \put(46,25){\line(1,0){1}}
    \put(47.4,25){\line(1,0){1}}
    \put(48.8,25){\line(1,0){1}}
    \put(50.2,25){\line(1,0){1}}
    \put(51.6,25){\line(1,0){1}}
    \put(53,25){\line(1,0){1}}
    \put(44.4,24,4){\line(1,1){1.2}}
    \put(44.4,25,6){\line(1,-1){1.2}}
    \put(54.4,24,4){\line(1,1){1.2}}
    \put(54.4,25,6){\line(1,-1){1.2}}
    \put(64.4,24,4){\line(1,1){1.2}}
    \put(64.4,25,6){\line(1,-1){1.2}}
    \put(65,25){\circle{2}}
    \put(56,25){\line(1,0){1}}
    \put(57.4,25){\line(1,0){1}}
    \put(58.8,25){\line(1,0){1}}
    \put(60.2,25){\line(1,0){1}}
    \put(61.6,25){\line(1,0){1}}
    \put(63,25){\line(1,0){1}}
  \end{picture}
\end{equation*}
\begin{equation*}
\begin{gathered}
   \Gamma_{123}=
   \begin{pmatrix}
      -1&-1&0\\
      -1&-1&-1\\
      0&-1&-1
   \end{pmatrix},\quad
  \mathbb{C}_{111}=\frac{1}{6}\left(\omega_{2}+\omega_{3}-2\omega_{1}\right),\quad
  \mathbb{C}_{222}=\frac{1}{6}\left(\omega_{1}+\omega_{3}-2\omega_{2}\right),\\
   \mathbb{C}_{333}=\frac{1}{6}\left(\omega_{1}+\omega_{2}-2\omega_{3}\right).
  \end{gathered}
\end{equation*}
Using these three-point blocks as well as the rules defined in section  \ref{walgebra} one can construct general solution.

\bibliographystyle{MyStyle} 
\bibliography{MyBib}
\end{document}